# Quasi-particle interference studies of quantum materials

Nurit Avraham, Jonathan Reiner, Abhay Kumar-Nayak, Noam Morali, Rajib Batabyal, Binghai Yan, Haim Beidenkopf

Condensed Matter Department, Weizmann Institute of Science, Rehovot 7610001, Israel

**Exotic electronic states are realized in novel quantum materials. This field was revolutionized by the topological classification of materials. Such compounds necessarily host unique states on their boundaries. Scanning tunneling microscopy studies of these surface states have provided a wealth of spectroscopic characterization, with the successful cooperation of *ab initio* calculations. The method of quasi-particle interference imaging proved to be particularly useful for probing the dispersion relation of the surface bands. Here we review how a variety of additional fundamental electronic properties can be probed via this method. We demonstrate how quasi-particle interference measurements entail mesoscopic size quantization and the electronic phase coherence in semi-conducting nanowires; helical spin protection and energy-momentum fluctuations in a topological insulator; and the structure of the Bloch wavefunction and the relative insusceptibility of topological electronic states to surface potential in a topological Weyl semimetal.**

Exotic electronic phases are realized and engineered in modern condensed matter materials, electronic systems and devices. One path for realization of such electronic systems is through the topological nature of their bulk band structure.[1-3] The electrons in those systems assume extraordinary properties including relativistic-like behavior, immunization from interaction with the environment and more. The topological phase can be induced by internal perturbations as strong spin-orbit interaction and the crystal field as is the case in many identified compounds. Alternatively, materials can be rendered topological by application of external perturbations such as strong magnetic fields in the case of the quantum Hall effect [4] or combination of magnetism and induced superconductivity in semiconducting nanowires with strong spin-orbit coupling which is thought to induce a topological superconducting phase.[5-11]

Here we survey our studies of the electronic properties of the unique states in quantum materials using the method of quasi-particle interference (QPI). In this method the scattering of the surface electrons due to surface imperfection, such as point impurities and step edges, leave a trace in the form of interference patterns in the local density of states (LDOS) that can be mapped through scanning tunneling microscopy (STM) and spectroscopic measurements. We demonstrate how a wealth of electronic properties can be extracted from such QPI patterns. In section 1 we exemplify the visualization of scattering processes from both point defects and step edges in Cu(111), and demonstrate their relation to the material band structure. In section 2 we investigate size quantization in semiconducting nanowires and identify a new energy regime in which hot-electrons surprisingly regain phase coherence with increasing energy. In section 3 we investigate the scattering properties of Dirac surface states of a topological insulator and their response to charged defects. In section 4 we study a topological Weyl semimetal and show how we use QPI measurement to resolve the structure of the wavefunction of the electronic surface states and hence to distinguish between coexisting trivial and topological surface bands.

## 1. Quasi particle interference (QPI)

The method of QPI have been used to study a variety of electronic systems including metallic surface states [12-14], various correlated states as high temperature superconductors [15,16] heavy fermion systems [17,18] and charge density waves[19], grapheme [20-22], and topological electronic systems as topological insulators [23,24], Weyl semimetals[25-27], and Majorana states [28]. We first demonstrate the application of the QPI method on the well-studied system of the two-dimensional Shockley surface states bound on the (111) surface of Cu single crystals.[12] The band structure of these states is known to be fairly free-electron like.[29] It has a parabolic dispersion, $E(k)=\hbar^2k^2/2m^*$, with an isotropic effective mass of $m^*=0.4m_e$. The bottom of the surface state band lies about 440 meV below the chemical potential. The extremely weak spin-orbit coupling in Cu leaves the surface band practically doubly degenerate.[30] We use this electronic system as a case study for demonstrating the principles of QPI method. The LDOS is given by

$$\rho(r, E) = \sum_k |\psi_k(r)|^2 \, \delta(E - \varepsilon_k),$$

Where $\psi_k(r)$ is the electronic wave function and $\varepsilon_k$ is its energy. We use scanning tunneling microscopy to spatially map the LDOS through the tunneling differential conductance, *dI/dV*. The typical LDOS, $\rho_0$, found on Cu(111) surface far from any scatterer is shown in the bottom panel of Fig.1a. The broad peak-like feature in the LDOS that sharply onsets 440 meV below the Fermi energy (above the dotted line) is the surface state contribution to the surface projected bulk LDOS.

## 1.1. Quasi-particle interference from point defects

We begin with examining the response of the surface electrons to a point-like scatterer located at $r_0$ and adds a potential $V(r) = V_0 \delta(r - r_0)$. A typical surface topography taken around such point-like scatterer on the Cu(111) surface is shown in Fig.1a. The location of the scatterer appears in the topography as a local dip which is surrounded by a circular ripple-like pattern. These circular ripples are the interference pattern of the scattered electrons embedded in the LDOS. They testify on the existence of a point scatterer such as a vacancy or an adatom, at the origin of the circular waves. The common adatom species on the surface of Cu are CO molecules. These form a potential barrier for the electrons tunneling from tip. Consequently, although topographically the molecule extends out from the surface, in order to maintain constant tunneling current under constant bias, the tip moves closer to the surface. The CO molecule is accordingly registered as a topographic intrusion rather than an extrusion.

The interference of incoming and scattered electronic surface states around the point impurity gives rise to the observed standing wave-like pattern. By measuring a *dI/dV* linecut across the impurity as a function of bias voltage one can directly image the evolution of these standing waves as a function of energy. The energy dispersion of the standing waves obtained around impunity on the surface of Cu(111) is shown in Fig.1b. It was measured along a line cut that crosses the impurity as marked by the blue line in Fig. 1a. Note, that the wavelength of these standing waves increases upon decreasing the energy until it diverges at ~ -440 meV which marks the bottom of the surface band. This dispersing wavelength signifies the dominant momentum transfer wavevector, $\vec{q} = \vec{k^i} - \vec{k^f}$, between initial and final states with the same energy $\varepsilon_{ki}=\varepsilon_{kf}$. In Cu(111) it corresponds to the backscattering channel $\vec{k^i} = -\vec{k^f}$, hence $\vec{q} = 2\vec{k^i}$. At the Femi energy this momentum transfer equals $\vec{q} = 2\vec{k_F}$. This can be obtained by solving explicitly the scattering problem with cylindrical symmetry imposed by the scatterer that yields to lowest order:

$$\rho(r,E) = \rho_0 \left\{ 1 + \frac{2}{\pi k_F r} \left[ \cos^2\left(k_F r - \frac{\pi}{4} + \eta_0\right) - \cos^2\left(k_F r - \frac{\pi}{4}\right) \right] \right\},$$

where $\eta_0$ is the phase shift affected by the detailed scattering potential.[12,15] For a $\delta$-like potential with $\eta_0 = \frac{\pi}{2}$ the modulated LDOS reduces to:

$$\rho(r,E) = \rho_0 \left\{ 1 - \frac{2\sin(2k_F r)}{\pi k_F r} \right\}.$$

The $\vec{q} = 2\vec{k_F}$ oscillatory modulation becomes apparent, as does the radial attenuation in the amplitude of the oscillations. This suppression originates from the spreading out of the standing wave pattern over the circumference that grows as $r$. The quantization of the dispersing band seen in Fig.2b results from the presence of an additional adjacent scattered (not imaged).

To extract the dispersion of the scattering wave vector $E(\vec{q})$, which for the simple surface band structure of Cu(111) is directly related to its energy dispersion $E(\vec{k})$, a Fourier transform is applied. The quadratic dispersing mode shown in Fig.1c was obtained by a Fourier transform of Fig.1b. It captures the exact dispersion of the surface band on Cu(111) surface.

### 1.2. Quasi-particle interference from step edges

Crystallographic step edges, as the one shown by the topography in Fig.1d present another type of scattering source with different geometry than the one presented by a point impurity. The atomically ordered step edge yields a translational invariant scattering potential along the step edge, $V(r) = V_0\Theta(x)$, where x is the distance from the step edge. As a result the impinging electrons undergo a unidirectional scattering perpendicular to the step edge such that $k_x^i = -k_x^f + q_x$ and $k_y^i = k_y^f$. In a material with isotropic dispersion as Cu summing over these scattering processes yields a LDOS given by

$$\rho(r, E) = \rho_0\{1 - J_0(2k_F x)\},$$

where $J_0$ is the zero-order Bessel function which asymptotically goes as $\cos(qx)/\sqrt{qx}$. [12,31] The resulting standing wave pattern along a line cut normal to the step edge (blue line in Fig. 1d) is shown in Fig.1e. Its Fourier transform, presented in Fig.1f, shows the energy dispersion $E(q_x)$ which remarkably agrees with the $E(q)$ obtained from scattering off a point impurity.

Finally, the bottom panel in Fig.1d presents a single energy cut of the standing wave pattern taken at 200 meV (dotted line). The expected oscillatory behavior with $x^{-1/2}$ decay is fitted to it (solid line). It becomes evident that the decay of the oscillation amplitude is stronger than that predicted. The reason for that is decoherence of the scattered electrons away from the step edge. The injected hot electrons lose their phase coherence due to interaction with the environment which facilitates their relaxation. These can be either interactions with the cold electrons occupying the Fermi sea or with phonons. In any case it leads to an exponential suppression of the standing wave amplitude as a function of the distance from the step edge

$$\rho(r, E) = \rho_0\{1 - e^{-x/L_\phi} J_0(2k_F x)\}$$

over a length scale defined as the phase coherence length, $L_\phi$. [32]

The above examples demonstrate that measuring the interference patterns electrons embed in the LDOS as they scatter off various crystallographic irregularities allows one to study the electronic dispersion, the scattering processes among it as well as the phase decoherence of the scattered electrons. In the following we demonstrate how all these aspects were used in our studies of electronic systems of distinct nature and dimensionality.

## 2. Quasi-particle interference study of semiconducting nanowires

Semiconducting nanowires have attracted vast scientific attention over the years.[33] More recently, nanowires with strong spin-orbit coupling, such as InAs and InSb, have gained particular attention due to the detection of zero-bias conductance peaks when rendered topologically superconducting by the combination of their spin-orbit interaction, induced superconductivity, and magnetization.[5-11] These possibly signify the existence Majorana end modes at the nanowire ends. However, this exciting observation is one out of many counter intuitive phenomena that were found in quasi-one dimensional electronic systems.[34] Some phenomena, such as spin-charge separation [35] and charge fractionalization [34], result directly from the Lutinger liquid properties of the system.[36] Other phenomena, such as perfect stability of hot holes [37], originate from the low dimensionality of the system.[38]

A major challenge in probing nanowires with surface probes such as STM is the native oxide layer that is formed when the wires are exposed to ambient conditions. An important exception is carbon nanotubes, in which the inert surface allows straightforward STM measurements, therefore serving as a rich platform to explore 1D electronic structure [39-41] and even Luttinger liquid behavior [42]. Semiconducting nanowires, on the other hand, have reactive surfaces that require employing aggressive means to remove the oxide layer, eventually damaging the pristine surfaces of the nanowires. For instance, small diameter silicon nanowires were treated with hydrofluoric acid that removes the oxide and passivate the surface with hydrogen. The hydrogen terminated surface showed atomic resolution and diameter-dependent energy gaps were measured [43]. III-V nanowires were studied after atomic hydrogen cleaning, showing crystalline surfaces and semiconducting spectrum [44-48]. This cleaning method did not suffice in order to resolve more subtle effects like the quantum confinement and QPI. Quantum confinement has been observed up to room temperature by Fermi level pinning at the bottom of the 1D subbands by Kelvin probe force microscopy of InAs nanowire [49].

We have resolved that technological challenge by constructing an ultra-high vacuum suitcase, which allows us to transfer the nanowires, under UHV conditions, from the molecular beam epitaxy (MBE) chamber, where the InAs nanowires are grown, to the STM chamber [50]. We first use it to transfer the Au substrates from the STM, where they are freshly prepared, to the MBE chamber. Then we harvest the nanowires mechanically by pressing the Au crystals against the nanowire substrate. As a result the nanowires break and attach to the Au crystal surface via Van der Waals forces. We bring the stamped Au crystals back into the UHV suitcase and carry them back to the STM chamber, where the sample with the best optically detected stamping mark is inserted into the cryogenic environment of the STM.

The outcome of this transfer procedure is seen in Fig.2. We search blindly via course motion for nanowires that lie on the Au substrate. A topographic encounter with two InAs nanowires that reside on the Au surface is shown in Fig.2a. The height of the nanowire from the substrate is about 70 nm, in agreement with height obtained from SEM images of the same batch. Such an object is a huge extrusion for the STM tip. Therefore, the side facet that we image represents the convolution between the nanowire side facet topography and the STM tip morphology. Accordingly we assume that only the top most facet of the nanowire, the Au substrate, and the height difference between them are true topographic representations. By zooming in to these top facets we atomically resolve the two possible terminations of a Wurzite InAs nanowire grown along the <0001> direction, which are the {11-20} and {10-10} side facets shown in Fig.2b and c, respectively. The insets show the correspondence between the topographic pattern image and the atomic structure. We indeed find very few impurities on the surface that were presumably added during the transfer procedure.

## 2.1. Quasi particle interference from point impurities in nanowires

Our all-UHV nanowire transfer method allows us not only to image the nanowire topography, but first and foremost to investigate spectroscopically the electronic states in it. The dI/dV spectrum measured on the facet imaged in the top panel of Fig.3a is shown in its bottom panel. At low energies we identify the semiconducting gap that terminates with the onset of the conduction band at about 100 meV below the Fermi energy. The conduction band appears as a series of peaks. They signify the thermally broadened van Hove singularities expected at the bottom of the quantized subbands.[51] An ab initio calculation of the dispersion of a Wurzite InAs nanowire of corresponding diameter is shown in Fig.3b. The orange line in Fig.3a is the integrated density of states derived from the calculated dispersion in Fig.3b. The good agreement between the measured and the calculated LDOS confirms the existence of the quasi-one dimensional subbands in the nanowire. This suggests that our InAs nanowires have 3-4 partially filled subbands that cross the Fermi energy with a typical inter-band energy spacing of a few tens of meV.

The quasi-one dimensional electrons scatter off the point impurities on the surface of the nanowire, giving rise to a complex quasiparticle interference pattern. It is exemplified in the dI/dV maps shown in Fig.3c, which were taken at three different energies. It shows the spatial fluctuations in the LDOS about the mean value. The strength of these fluctuations is less than 10% of the total LDOS, signifying that the point impurities are weak scatterers of the quasi-one dimensional electrons. Spatial integration of the LDOS, shown in Fig.3d, yields the expected series of van Hove singularities. Remarkably, Fourier transformation analysis finds a series of dispersing modes marked by dotted lines. The bottom of each mode corresponds to a van Hove singularity detected in the average LDOS. This correspondence serves as an important consistency check that the modes visualized indeed signify the quantized subbands in the nanowire. The extracted effective mass of these subbands is $m^*=0.05\pm0.01m_e$, in agreement with the effective mass of the conduction band of bulk InAs.[52]

## 2.2. Quasi particle interference from the nanowire end

We now turn to discuss the scattering off the strongest possible scatterer in nanowire – the nanowire end. The topography of the end attached to the Au droplet that catalyzed the growth is shown in Fig.4a. We measure a dI/dV line cut along the yellow dashed line that yields the LDOS image in Fig.4b. Far from the nanowire end, on the right side of the image, we clearly resolve the non-dispersing van Hove singularities. Close to the nanowire end we find, on top of the non-dispersive peaks, a clear dispersing standing wave pattern that emanates from the nanowire end (left hand side of the image). The two distinct behaviors seem to contradict. On the one hand, the van Hove singularities suggest that many subbands occupy the imaged energy window. On the other hand, we find only a single coherent standing wave pattern suggesting that only one of the subbands gives rise to QPI pattern. The latter is clearly captured by the Fourier transformation of the line cut map, given in Fig.4c. It shows a single strong dispersing mode (marked in dashed line) whose dispersion agrees perfectly with that of the calculated lowest subband of the conduction band in Fig.3b. We also note that taking the Fourier transformation along the quantized direction that lies across the nanowire axis, as shown in Fig.4d, yields no detectable dispersion as it should.

Intriguingly, we find that the resolution of this conflict lies in the markedly different phase coherence of the lowest quantized subbands and that of higher ones. As sketched in Fig.4e, hot electrons injected to higher quantized subbands can relax via inter-band process by exciting a corresponding inter-band electron-hole pair from the Fermi-sea. This is a two-body process requiring no momentum transfer with broad phase space of available states. Accordingly, the rate of relaxation of these hot-electrons is high, and their phase coherence is short. In contrast, hot electrons injected to the lowest quantized subband can only relax via inta-band processes that were shown to involve excitation of two electron-hole pairs to conserve both energy and momentum.[53,54] The phase space of available states is much narrower and the amplitude of such three-body process is much weaker. Consequently, hot electrons from the lowest quantized subband will relax much more slowly, and their phase coherence length will be much longer. This exemplifies the sensitivity of the QPI method to the phase coherence of the electrons.

In the following we use the QPI method to resolve the energy dependence of the phase coherence length and thus find a new high energy regime with extended coherence of the one-dimensional electrons. One dimension provides the optimal setup for investigating the phase coherence length. As we've shown in section 1, in higher dimensions the amplitude of the QPI pattern decays radially as a power low because the wave function spreads out. In the quasi-one dimensional limit the nanowire confinement collimate the wave function of the electrons. As a result, in the absence of decoherence the QPI pattern does not decay and is simply given by

$$\rho(r, E) = \rho_0\{1 - \sin(2k_F x)\}.$$

Therefore, any trace of decay indicates a phase decoherence of the electronic states.

We begin by eliminating from Fig.4b all the features that do not oscillate with the frequency we identify by Fourier analysis in Fig.4c (primarily the van Hove singularities). This yields the intensity map shown in Fig.5a. It already captures some clear trends. At the lowest energies, in the hot holes sector, we find extended phase coherence that grows towards the Fermi energy in full agreement with previous studies. Above the Fermi energy the Fermi coherence length rapidly shortens with energy as the phase space of electrons to interact with grows. However, above a certain energy threshold of about 70 meV the phase coherence length saturates and then revives. Remarkably, at the highest energies measured, the oscillatory modulation of the LDOS extends throughout the field of view without any apparent decay. We quantify this behavior by fitting an exponentially suppressed sinusoidal profile $\delta\rho(r,E) = \rho_0 e^{-x/L_\phi} \sin(2k_F x)$, as demonstrated in Fig.5b. We plot the extracted energy dependence of the phase coherence length $L_\phi(E)$ in Fig.5c, which clearly shows a high energy regime in which the phase coherence length revives.

The physical origin of this revival lies in the combination of two effects. First, confining potential provided by the nanowire, with a diameter $d$ affects the nature of the Coulomb interaction between electrons. It leads to the suppression of momentum transfers that are much higher than the inverse-diameter of the nanowire, $q \gg 1/d$. As a result, above a certain energy threshold the interaction of the hot electrons with cold electrons deep in the Fermi-sea involves exceedingly large momentum transfer and is therefore suppressed. This cut off thus suppresses the growth in phase space of cold electrons to effectively interact with. The second effect results from the velocity mismatch between that of the injected hot electrons and the Fermi velocity of the cold ones. The higher the energy of the hot electrons the more significant the velocity mismatch is. As the mismatch grows, the hot electrons traverse the cold ones too fast to interact effectively giving rise to a decrease in the interaction strength with increasing energy. These two ingredients are captured in a designated model shown by the solid line in Fig.5c. It agrees well with the non-monotonic trend we observe in our experiment. Furthermore, it predicts that the revival in the phase coherence will increase with increasing energy until a prefect decoupling of the ultra-hot electrons from the cold Fermi sea is obtained. This trend will be cut in realistic setups by other processes, such as the finite bandwidth of the quasi-one-dimensional band.

## 3. Quasiparticle interference study of topological insulators

Topological insulators are bulk insulators that host 2D Dirac fermions on their surfaces. These unique surface states result from an odd number of band inversions in their bulk band structure.[1-4] Such reordering of the valence and conduction bands, typically in response to strong spin-orbit coupling, does not land any spectacular bulk phenomena, however, it assures the formation of relativistic-like Dirac states on all the boundaries of the samples. These topological surface states exhibit a helical spin texture due to their strong spin orbit interaction. Topological insulators have found realization in a variety of compounds over the years.[55,56]

STM in general and QPI in particular, had a central role in the study of topological phases of matter. The first QPI study of this class of materials was on the first identified three-dimensional strong topological insulator $Bi_{1-x}Sb_x$.[23] By identifying all possible scattering wave vectors among the rather complex band structure of this alloy the absence of interference patterns originating from backscattering, or more precisely between states with orthogonal spin projections was detected. This was the first demonstration of topological protection against backscattering by the helical spin-texture of the topological surface states. Imaging of the Dirac band structure and protection against backscattering were later demonstrated in the canonical topological insulators Bi2Se3 and Bi2Te3 [24,57,58]. Even though direct backscattering is forbidden in the helical surface states of these materials, oblique scattering is still allowed. As a result the QPI pattern is altered rather than completely eliminated.[59,60]. The spin protection will also affect the power-lay decay in the amplitude of the QPI patterns away from the impurities [61]. The Dirac dispersion was further probed through Landau levels spectroscopy of the topological bands [62-64].

Here we present our QPI study of the response of the topological Dirac surface states to disorder.[65] The surface topography of a $Bi_2Se_3$ single crystal doped with nominal 2.5% of Mn atoms is shown in Fig.6a. The Mn atoms substitute Bi and accordingly reside at least one monolayer below the Se layer terminating the quintuple unit cell. As a result, they alter the appearance of the three nearest neighboring Se atoms on the top most surface.

### 3.1. Quasi particle interference of Helical states

The Mn atoms act as strong scatterers for the surface electrons. We indeed find a dispersing QPI pattern in the dI/dV mappings shown at two different energies in Fig.6b and c. The dispersion of the surface states in $Bi_2Te_3$ changes from a circularly symmetric Dirac-like dispersion close to the Dirac node to a hexagonally warped dispersion at higher energies,[66-68] as shown in Fig.6d. The Fourier transformation of the dI/dV maps also exhibit a transition from circularly symmetric QPI close to the Dirac node (Fig.6g) to hexagonally symmetric at higher energies (Fig.6f).

To identify the origin of the different QPI patterns we calculate the joint density of states (JDOS)

$$JDOS(q,E) = \int dk\, \rho_k \langle \sigma_k | \sigma_{k+q} \rangle \rho_{k+q},$$

where $\rho$ is the calculated dispersion and $\sigma$ is the calculated spin texture.[23] The effect of the spin protection is clearly observed in the JDOS of the high-energy warped band. The star-like band structure gives rise to quasi-nested scattering conditions with an elaborate spin texture that meanders in and out of plane.[69,70] The resulting JDOS is shown in the top panels of Fig.6e, where on the left the spin projection is neglected and on the right it is accounted for. The spin texture associated with the warped band suppresses the QPI peaks along the Γ-K direction while leaving those along Γ-M quite unchanged. Remarkably, in the measurement we indeed find QPI patterns along the Γ-M direction alone (Fig.6b). This demonstrates the spin protection against scattering.

At low energies close to the Dirac node the dispersion is linear and fairly circularly symmetric with an in-plane helical spin texture associated with it. This Dirac-like dispersion eliminates direct backscattering,[23] and by this has a far-reaching effect on the conductivity as it prevents weak localization.[1-4] However, protection from backscattering has a rather subtle effect on the QPI pattern.[70] Comparing the calculated JDOS with and without accounting for the helical spin protection (Fig.6e, bottom right and left panels, respectively) one finds in both a $q < 2k_F$ disc of available scattering channels. The main qualitative difference introduced by the helical spin texture is the elimination of a $q = 2k_F$ ring, at the margin of the $q < 2k_F$ disc, which is associated with direct backscattering. Interestingly, the measured low energy QPI pattern demonstrated in Fig.6c does have a bounding ring. We note, that similar $q = 2k_F$ scattering rings were detected in QPI of other Dirac systems as the surface state in the topological crystalline insulator SnTe,[71] as well as in graphene.[21] The origin of this unexpected behavior in yet unclear, but its reoccurrence may suggest that the protection from backscattering introduced by the helical spin-texture is not complete.

**3.2. Quasi particle interference visualization of momentum fluctuations**

The QPI patterns discussed above are not the sole origin of the fluctuations observed in Fig.6b and even more clearly in Fig.6c. We identify an additional non-dispersing inhomogeneity in the LDOS. Plotting the dI/dV spectrum along a linecut across the field of view, shown in Fig.6h, we find that the whole spectrum shifts rigidly in energy. The charged Mn dopants introduce an underlying electrostatic potential. This potential is screened by the electrons, and accordingly induces fluctuations in the LDOS. The Dirac spectrum fluctuates across the surface by $\Delta E \approx 20$ meV, demonstrated by the extracted profile in Fig.6i. As a result the energy of the Dirac node can no longer be globally defined as it fluctuates spatially.

We use the method of QPI to image the corresponding spatial fluctuations in momentum. We split the mapped field of view into regions with LDOS higher and lower than the average value, as shown in Fig.7a. The typical length scale of the charge puddles is $l \approx 200$ nm. We then Fourier transform each of the sub-region separately. Remarkably, instead of the original broad QPI peak (grey line in Fig.7b) we find two narrower QPI peaks (blue and red lines in Fig.7b) shifted from one another by $\Delta q = 0.01$ Å$^{-1}$. This verifies that the surface electrons modify their momentum as they traverse the underlying electrostatic potential.

We repeat this analysis at different energies and find a fairly energy-independent momentum shift, as shown in Fig.7c. The relation between the momentum and energy shifts agrees well with the Fermi velocity extracted from the QPI dispersion $\frac{\Delta E}{\hbar \Delta q} \approx v_F = 1.3\ eV\ Å$. Below the momentum scale given by $q \approx 2\pi/l$ we can no longer resolve any QPI patterns. Once the electronic wavelength, that diverges on approaching the energy of the Dirac node, becomes larger than the typical length scale of the charge puddles, the momentum ceases to be a well-defined quantum number. In this low-energy range the momentum fluctuations become larger than its mean value. A similar phenomenon was detected before in graphene when exfoliated on $SiO_2$.[21,72] It was later resolved by replacing the $SiO_2$ with cleaner substrates as hexagonal boron nitride [73] or altogether suspending it. [74]

### 4. Quasi particle interference study of topological semimetals

A Weyl semimetal forms in a material in which the bulk is gapped except for at an even number of points in the Brillion zone, in which the bands touch, that are called Weyl nodes. Near the touching points the electrons have a three dimensional Dirac dispersion and a well-defined chirality. A Weyl point can be considered as a source or sink of Berry flux, depending on its associated chirality. Consequently, for 3D samples, on those surfaces where bulk Weyl points of opposite chiralities do not project to the same point in the surface Brillouin zone, a topological surface Fermi-arc band will form. This surface band connects two surface projections of Weyl nodes with opposite chirality. Accordingly Fermi-arc surface bands seem to have an open contour.[75-77] This unique surface-bulk connectivity gives rise to various unique electromagnetic effects.[78-83]

### 4.1. Quasi particle interference from point impurities in TaAs

The QPI method have been used to study the existence of Fermi-arc states and their unique connectivity to the bulk Weyl cones in several material systems. These include TaAs [25,26], that we discuss here, and NbP [27]. The type two Weyl semimetal MoTe2, in which the Weyl cone dispersion is tilted such that the electron and hole cones overlap in energy, was also probed [84] and the related Dirac semimetal $Cd_3As_2$.was studied using Landau level spectroscopy. [85] We have used the method of QPI to uniquely characterize the structure of the wave function of the topological Fermi-arc surface states in the Weyl semimetal TaAs.[26] We further used it to distinguish the Fermi arcs from non-topological surface states that coexist on the sample surface. TaAs is a Weyl semimetal with a total of 24 bulk Weyl nodes which on the (001) surface project to 16 surface Weyl nodes with 8 Fermi arcs connecting them shown in DFT calculation in Fig.8a.[86-90] In addition there are trivial surface states induced by the dangling bonds on the exposed (001) surface.[27] This richness poses both a challenge of distinguishing the topological bands from the trivial ones but also an opportunity to compare the properties of the two kinds.[25] The topography of the cleaved surface is shown in Fig.8b. We find a perfectly ordered square lattice (zoomed in image shown in the inset) with a low concentration of atom vacancies. By comparing the measured dI/dV spectrum over that surface, shown by solid line in Fig.8c, to the calculated one for Ta versus As (red versus blue dashed lines, respectively) termination we identify the cleaved surface as As terminated. Accordingly the deficiencies are As vacancies. These scatter the surface electrons and give rise to the standing wave pattern captured by the dI/dV map in Fig.8d.

Fourier analysis, presented in Fig.8e, resolves the elaborate structure of scattering wave vectors that participate in the formation of that standing wave pattern. The brightest spots, marked with $\Gamma_x$ and $\Gamma_y$, are the atomic Bragg peaks. Based on their location we can divide the QPI pattern to scattering wavevectors shorter than the Brillouin zone (within the central dashed square) and scattering wavevectors which are larger than the Brillouin zone.

The JDOS calculated based on the band structure shown in Fig.8a is given in Fig.8f. The blue, yellow and green patterns correspond to scattering within the ellipse-like band, the bow-tie like band, and among them, respectively (as marked by corresponding colored arrows in Fig.8a). Accordingly, these are all QPI patterns from scatterings among trivial bands. The red QPI patterns in Fig.8f involve scattering with a Fermi arc surface band. This includes intra-Fermi-arc scatterings (Q3), inter-Fermi-arc scatterings (Q2) and scattering between a Fermi-arc and a trivial band (Q1, Q4). Among all scattering processes in Fig.8e we identify only the leaf-like pattern that peaks beyond the ellipse-like pattern, given in greater detail and compared to calculated JDOS in Fig.8g, as one involving Fermi-arcs (Q1).

The JDOS calculation captures only the scattering processes of wave-vectors shorter than the size of the Brillouin zone. Clearly, to account for higher scattering wave-vectors higher Brillouin zones have to be considered. Those higher Brillouin zones are a direct consequence of the periodicity of the crystal as manifested by the Bloch wavefunction:

$$\psi_k(r) = \sum_G c_{k,G} e^{i(k+G)\cdot r},$$

where $G$ is a reciprocal wave-vector. Accordingly, the translational invariant LDOS can be written as:

$$\rho(r, E) = \sum_{G,G'} A_{G,G'} e^{i(G-G')\cdot r},$$

$$A_{G,G'} = \sum_k C_{k,G}^* C_{k,G'} \delta(E - E_k).$$

This means that even in the absence of a scatterer the LDOS will be modulated by $g = G - G'$ whenever more than a single dominant Bloch coefficient, $C_G^k$, exists. This is indeed observed in the modulated LDOS in the inset of Fig.8d. On the same footing the JODS will now assume the form:

$$JDOS(q, E) = \sum_{G,G'} \int dk\, \rho_{k+G} \langle \sigma_{k+G} | \sigma_{k+q+G'} \rangle \rho_{k+q+G'},$$

which will replicate QPI features contained within the first Brillouin zone to higher ones whenever several dominant Bloch amplitudes, $C_G^k$, appear.[14]

Free particles can have pure plane-wave like behavior. However, electrons within periodically ordered material will be susceptible to some extent to the underlying potential imposed by the crystal. This will necessarily render higher Bloch coefficients to be non-vanishing. The relative strength of the different Bloch coefficients determines the structure of the electronic wave function within the unit cell. It can be predicted in ab initio calculation and measured by QPI. Representative calculated Bloch coefficients of the ellipse- and bowtie-like bands are given in Fig.9a and b, respectively. We indeed find an anisotropic structure which corresponds to the replications of QPI patterns found in experiment (Fig.8e). However, the topological Fermi arc states, in contrast to dangling bond states, result from the bulk topology rather than the surface potential. It raises the question to what extent are the topological surface states susceptible to the underlying crystalline structure. Calculation of the Bloch structure of the Fermi arc wave function, shown in Fig.9c, indeed finds a single dominant coefficient. This intriguingly suggests that Fermi-arc states are plane-wave like. Accordingly they will not be replicated to higher Bragg peaks in QPI. We indeed do not seem to detect any replication of the leaf-like structure, associated with scattering from a Fermi-arc state, to higher Bragg peak.

With this in mind, we perform a novel analysis on the QPI data. We subtract the QPI ellipse pattern that appears around $\Gamma_{\pm Y}$ from the QPI pattern that appears about q=0. By doing so we are indeed able to eliminate the ellipse-like QPI pattern, as seen in Fig.9d. In this procedure the leaf-like QPI patterns in which Fermi-arcs are involved are indeed hardly changed, signifying that this QPI pattern is indeed not replicated. Remarkably, once the ellipse-like pattern is eliminated we find a residual curly QPI structure that fits well the scattering pattern of the second Fermi-arc surface band along $\Gamma$-Y (see JDOS calculation in Fig.9e). This demonstrates that generally QPI patterns involving trivial bands are replicated according to the structure of their Bloch wave-function. The topological Fermi-arc states are found to be remarkably unsusceptible to the underlying crystal structure and be well approximated by a pure plan-wave like wave function.

### 4.2. Quasi particle interference from step edges in TaAs

Additional information on the structure of the wave function of the Fermi arc bands we obtain from their scattering properties off crystallographic step edges. A topographic image of such a step edge is shown in Fig.10a. The step edge is oriented 49º relative to the crystal axis and therefore scatters approximately along the $\Gamma$-M direction. The scattered electrons give rise to an intricate QPI pattern, shown in Fig.10b. It comprises dispersing features as well as atomic modulation, highlighted by the inset. We therefore separate the dI/dV linecut into two subsets – the dI/dV measured on the As atoms and in the valleys between them where the topmost Ta atoms reside (Fig.10c and d, respectively). Each of these subsets displays a distinct dispersing QPI pattern.

Indeed, Fourier transformation of each of them, given in Fig.10e and f, displays two distinct sets of dispersing scattering modes. On the As surface layer, presented in Fig.10e, we find the dispersion that corresponds to the ellipse-like QPI pattern of Fig.8e. Calculation indeed verifies that the ellipse band results from the As dangling bonds and is accordingly highly localized on the topmost As monolayer, as shown by the calculated wave function distribution at the inset. In contrast, the QPI modes that originate from the LDOS in between the topmost As atoms, shown in Fig.10f, finds a completely distinct dispersing modes. Comparison with the calculated JDOS, shown in Fig.10g, identifies them with scattering processes within the Fermi-arc located on the $\Gamma$-Y direction. They both disperse towards the energy and momentum at which the surface projection of the bulk Weyl node resides. The wave function distribution of that Fermi arc band, presented at the inset of Fig.10f, confirms that this topological band resides predominantly on the Ta sites and penetrates deeper into the bulk. Indeed, the bulk Weyl cones as well as the surface Fermi-arcs induced by them are derived mainly from the Ta orbitals.

## 5. Discussion

We have demonstrated that the method of QPI entails a wealth of properties that go beyond visualization of the material dispersion. Its origin in quantum interference between the incoming and scattered electrons provides direct access to the electronic phase coherence. Its direct relation with scattering processes allows to investigate this fundamental mechanisms as we did with immune topological surface states in a topological insulator. Spatial resolution further allows to map the electrostatic screening. With subatomic resolution the detailed structure of the electronic wavefunction can be mapped and analyzed. We show an interleaved QPI pattern where different bands reside of distinct atomic sites. We further detect the structure of the Bloch wave function. The latter allows us to employ a novel analysis, in which we deduct QPI replications from Bragg peaks of different orders from one another to uncover the distinct structure of the Bloch wave functions. Yet, we stress that in other contexts different operations can be applied on the various Bloch replications to gain access to particular properties of the Bloch wave function.[91,92]

**Acknowledgements**

We acknowledge Ali Yazdani, Robert J. Cava, Claudia Felser, Hadas Shtrikman, Ilya Drozdov, Jungpil Seo, Pedram Roushan, Ion Cosma Fulga, Jung-Hyun Kang, Torsten Karzig, Yan Sun, Marcus Schmidt, Ady Stern that participated in this research and made it possible. H.B. acknowledges support from the European Research Council (ERC) (Starter Grant no. 678702, "TOPO-NW"), the United States–Israel Binational Science Foundation (BSF), the Israeli Science Foundation (ISF), the GIF - the German-Israeli Foundation for Scientific Research and Development and the Minerva Foundation. B.Y. acknowledges the support by a research grant from the Benoziyo Endowment Fund for the Advancement of Science. H.B. N.A. and B.Y. acknowledge funding by the German-Israeli Foundation for Scientific Research and Development (GIF Grant No. 308 I-1364-303.7/2016).

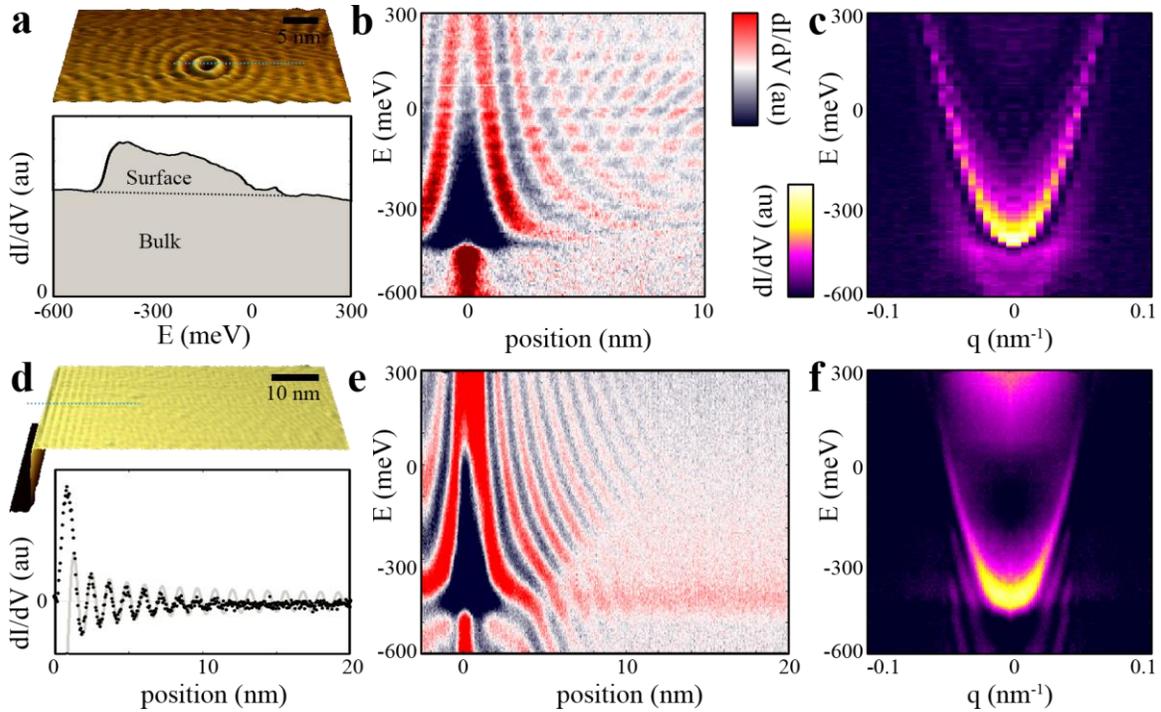

Figure 1. QPI in Cu(111). a) Upper panel: low bias topographic image of a CO molecule on Cu(111) surface. Lower panel: dI/dV spectrum away from scatterers showing the bulk and surface states contribution to the LDOS. b) Dispersing QPI pattern at a line cut across the CO molecule (along the dotted line in a). c) Fourier transformation of b shows the surface states dispersion E(q). d) Upper panel: low bias topographic image of a crystallographic step edge on Cu(111) surface. Lower panel: dI/dV linecut at 200 meV away from the step edge (along dotted line) fitted with appropriate Bessel function. The measured decay is faster due to decoherence. E) Dispersing QPI pattern at a line cut across step edge (along the dotted line in d). f) Fourier transformation of e shows the surface states dispersion E(q).

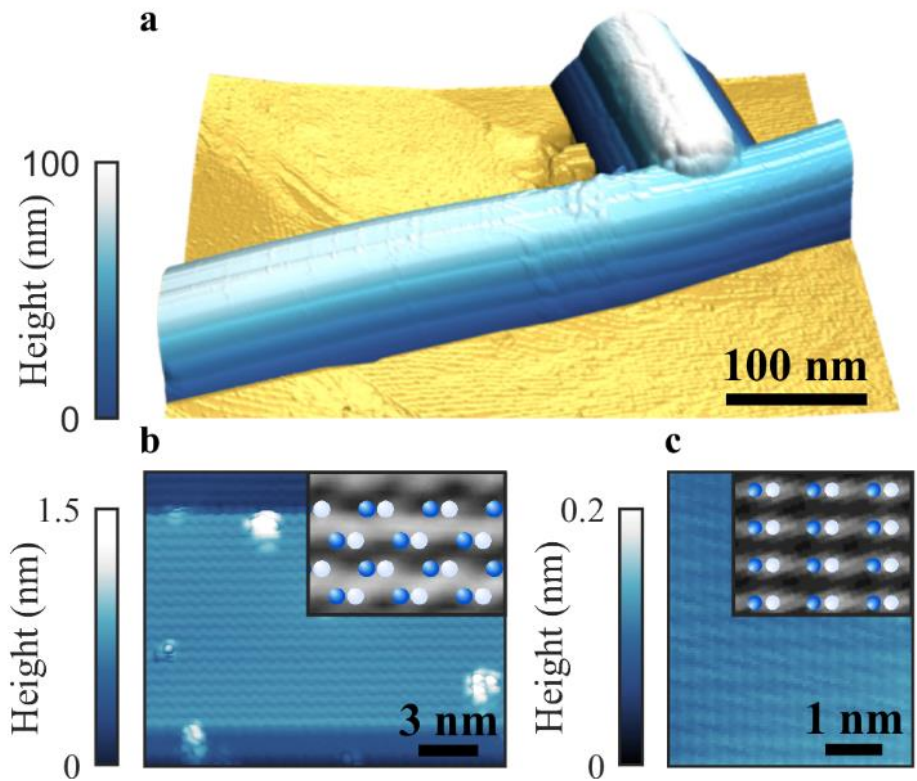

Figure 2. Nanowire topography. a) Topographic image of two InAs nanowires on top of Au(111) substrate. b,c) Atomically resolved top facets. Insets shows agreement with atom arrangement of the {11-20} facet (b) and {10-10} facet (c). Adapted with permission.[50] Copyright 2017, American Physical Society

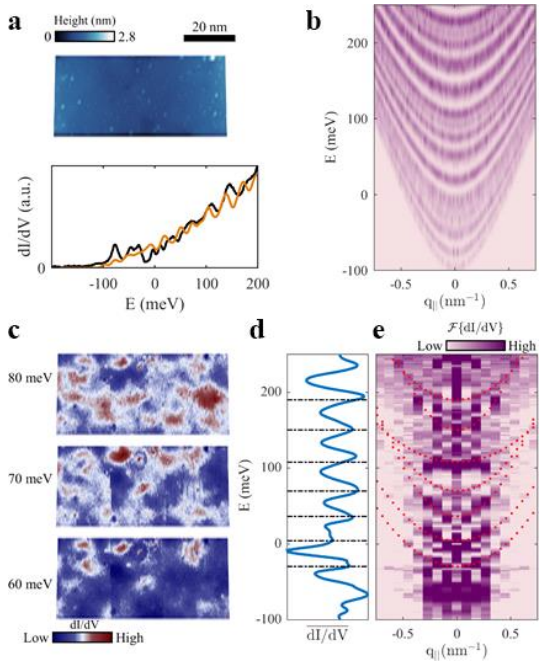

Figure 3. Quantization. a) dI/dV spectrum (bottom panel) taken over the nanowire (topography in top panel) showing a series of van Hove resonances that onsets 100 meV below the Fermi energy. It agrees with ab initio calculation of the spectrum of a nanowire with similar parameters. b) Calculation of the spectrum of Wurtzite InAs nanowire quantized into subbands. c) dI/dV mapping of the LDOS across the nanowire segment imaged in a. d) Van Hove resonances seen in the spatially averaged LDOS. e) QPI visualization of the dispersing quantized subbands by Fourier transformation of the dI/dV maps. The dotted lines are parabolic fits. Their minima are consistent with energies of the van Hove resonances. Adapted with permission.[50] Copyright 2017, American Physical Society

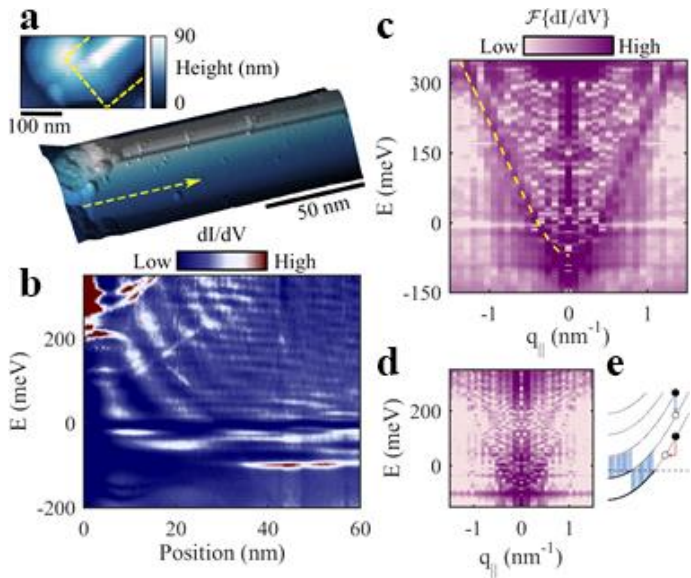

Figure 4. QPI at the nanowire end. a) Topographic image of the end of a nanowire. Inset shows the Au droplet that catalyzed the nanowire growth. b) dI/dV linecut measured away from the nanowire end (along dashed arrow in a) showing a dispersing QPI mode. c) Fourier analysis of b (along the nanowire axis) showing the dispersion of the lowest subband. d) Fourier analysis across the nanowire does not capture any signature of QPI. e) The origin of the strong QPI signal of the lowest subband is its extended phase coherence. Adapted with permission.[50] Copyright 2017, American Physical Society

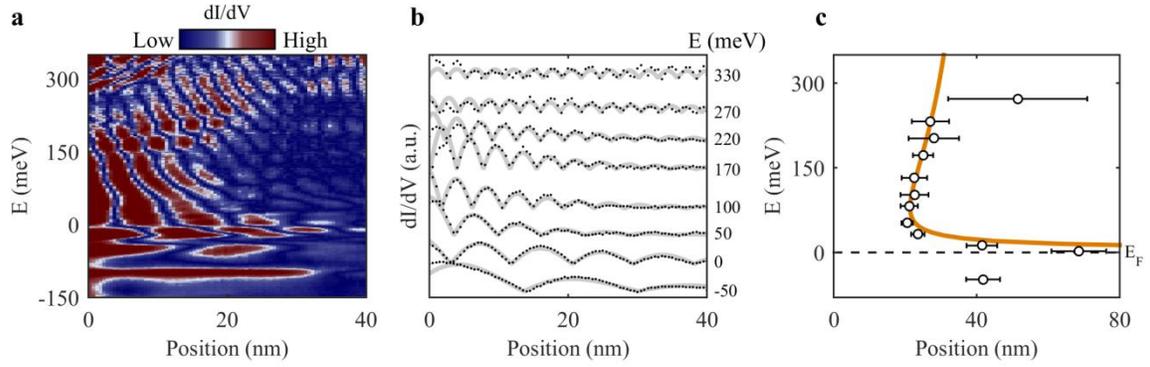

Figure 5. Revival of phase coherence length of quasi-one dimensional hot electron. a) absolute value of the LDOS in Fig.4b after filtering out the non-modulated features. b) exponentially suppressed sinusoidal fits to the filtered LDOS. c) Energy dependence of the phase coherence length shows a non-monotonic behavior above about 70 meV. The evolution of the phase coherence length is well captured by our model (solid line). Adapted with permission.[50] Copyright 2017, American Physical Society

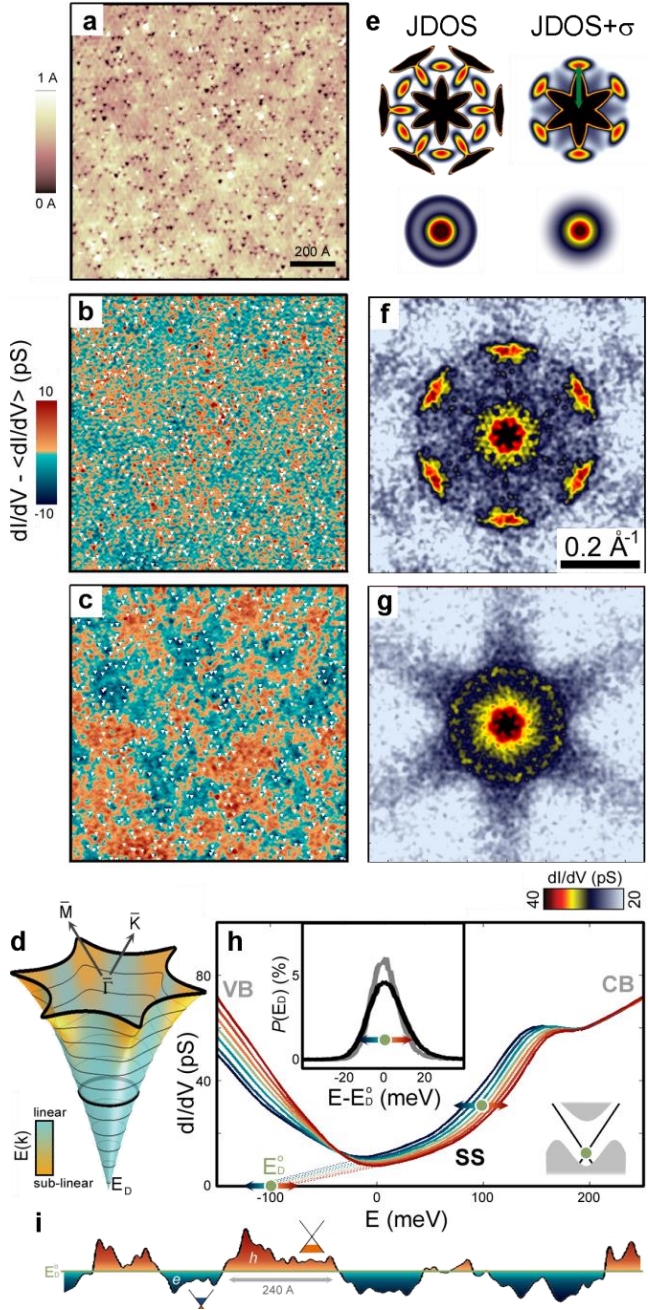

Figure 6. QPI in Bi2-xMnxSe3 and energy fluctuations of the spectrum a) Topographic image of the cleaved surface. b) dI/dV map at the energy of 100meV. c) dI/dV at the energy of 300meV. d) Surface band structure. e) JDOS calculated without (left) and with (right) the helical spin texture and for hexagonally warped dispersion (top) and linear Dirac dispersion (bottom). f) Fourier transform of b. g) Fourier transform of c. h) dI/dV spectrum at equally spaced points along a linecut taken across the mapped region. The inset shows the distribution of Dirac node energies across the images surface. i) A linecut across the surface showing the fluctuations of the Dirac node energy and the electron-

hole charge puddles it induces. Adapted with permission.[65] Copyright 2011, Macmillan Publishers Limited.

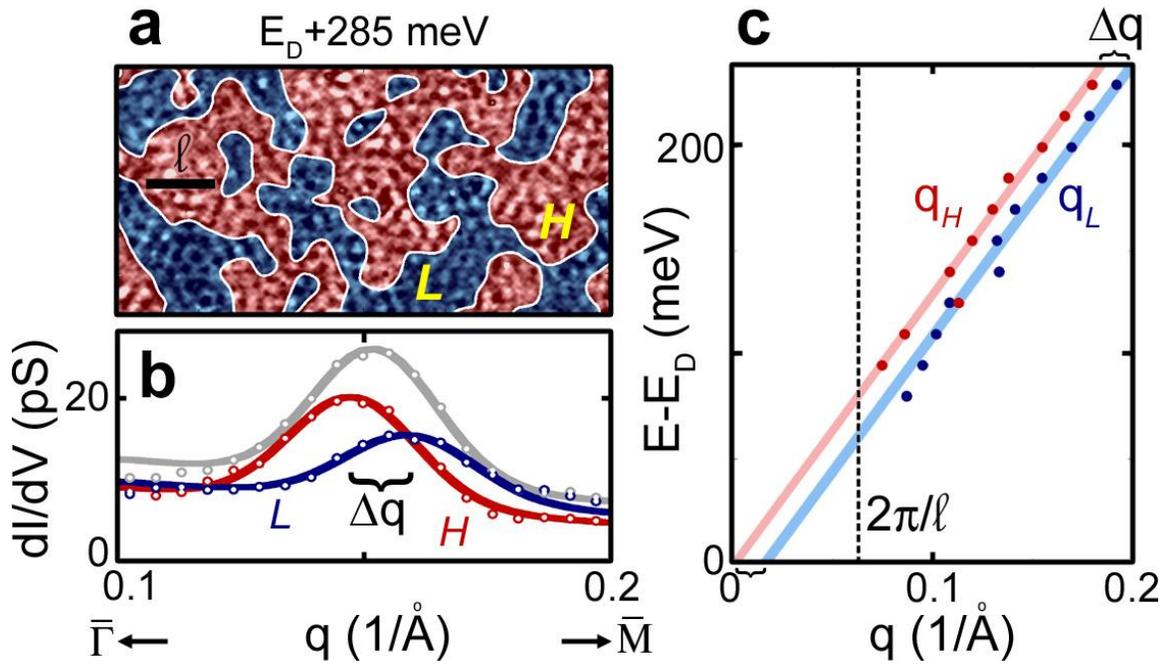

Figure 7. QPI visualization of the momentum fluctuations. a) Separation of the dI/dV map into regions with a LDOS higher and lower than its mean value (red and blue, respectively). b) Cut along G-M of the Fourier transform of the whole map in a (grey line) and of the red and blue regions separately (red and blue lines, respectively). c) The energy dispersion of the maximal momenta of the G-M QPI peaks of the regions with high and low LDOS (red and blue, respectively). Adapted with permission.[65] Copyright 2011, Macmillan Publishers Limited.

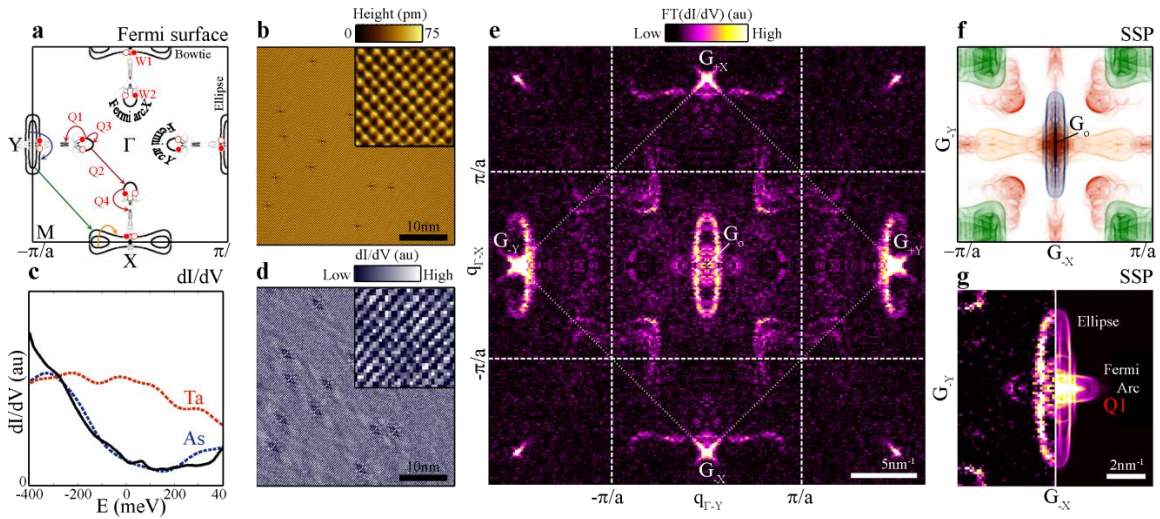

Figure 8. QPI in TaAs. a) The calculated Fermi surface of TaAs (110) surface comprising of both trivial bands and topological Fermi-arc states. The arrows mark possible scattering processes with the surface band structure. b) Topographic image of the surface. c) comparison of the measured dI/dV spectrum (solid line) with calculated spectrum of As versus Ta terminated surfaces (blue versus red lines, respectively) suggests the cleave exposes the As layer. d) dI/dV map at the Fermi energy finds complex QPI patterns around each As vacancy imaged in b. e) Two dimensional Fourier transformation of d. f) calculated JDOS based on a. g) The leaf-like QPI pattern that peaks beyond the central ellipse corresponds to scattering between the Fermi arc along Γ-Y and an adjacent trivial band. Adapted with permission.[26] Copyright 2016, American Association for the Advancement of Science.

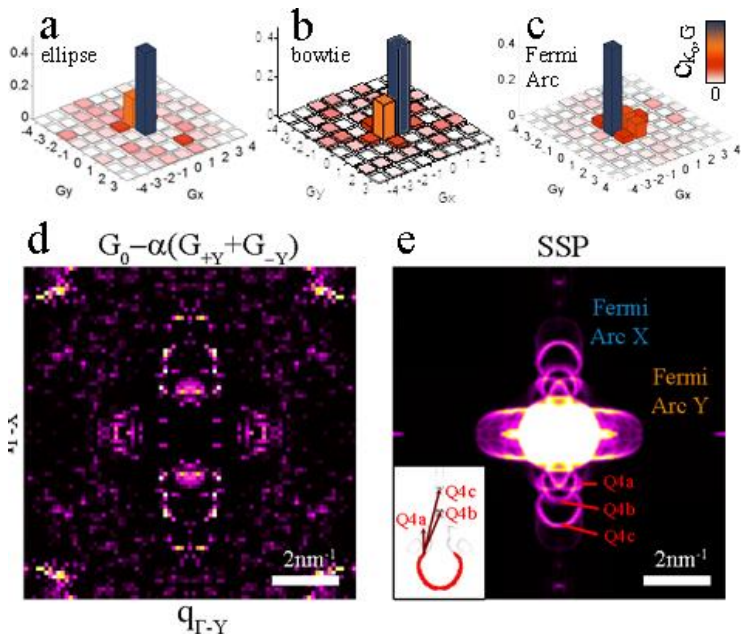

Figure 9. Visualizing the Bloch wave function in QPI. a-c) Calculated Bloch coefficients of three different bands. The ellipse and bowtie bands have a complex anisotropic Bloch structure while the Fermi-arc band is rather plane-wave-like with a single dominant Bloch coefficient. d) Subtracting the ellipse-like QPI of the Γ-Y Bragg peak from that at q=0 reveals previously buried QPI pattern. e) Calculated JDOS about q=0 in the absence of ellipse and bowtie bands. Adapted with permission.[26] Copyright 2016, American Association for the Advancement of Science.

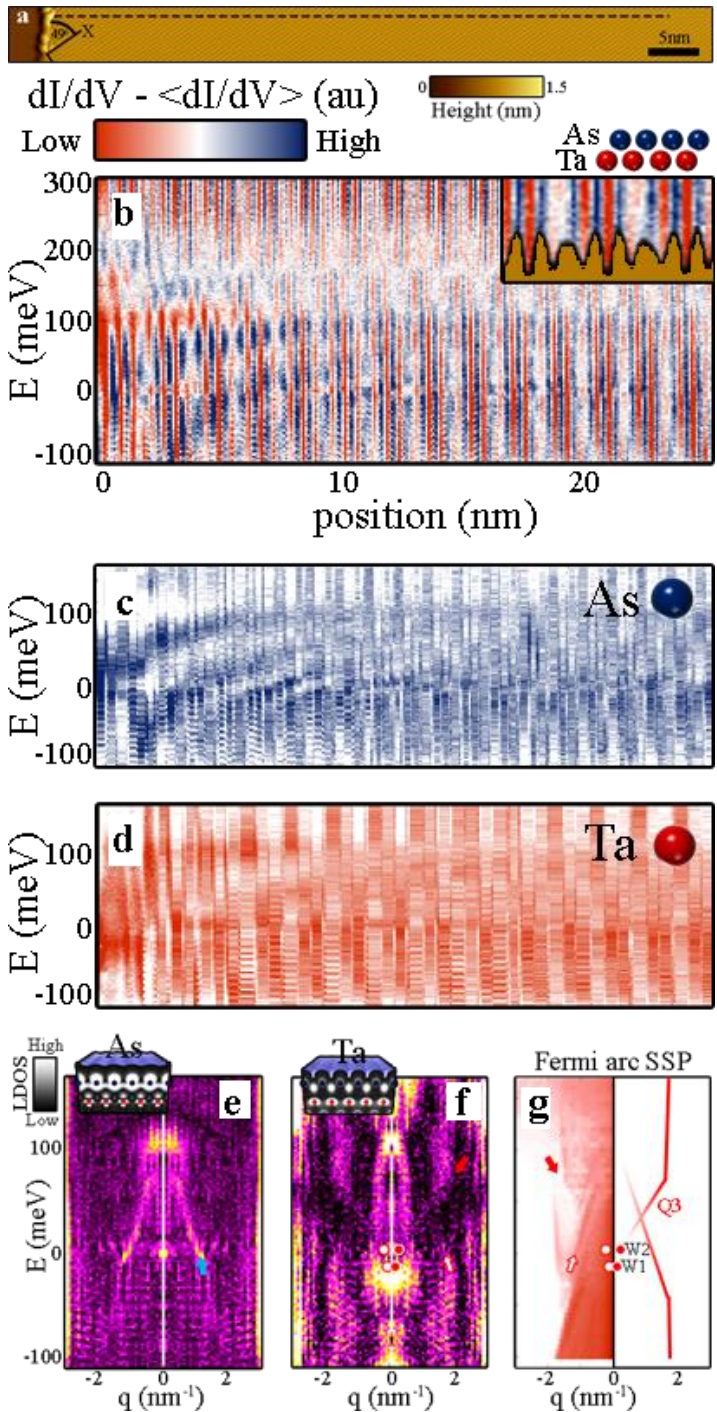

Figure 10. Distinct atomic origin of trivial and topological surface bands. a) Topographic image of a crystallographic step edge. b) dI/dV linecut measured normal to the step-edge (along dashed line in a). c,d) dI/dV linecut taken only on As (c) or Ta (d) atomic sites. e) Fourier analysis of c finds the dispersion of the trivial ellipse band derived from the As dangling bonds. Inset shows that the calculated wave function distribution of the ellipse

band is indeed highly localized on top-most As layer. f) The QPI mode on the Ta sites differs from that seen on the As sites. Inset shows the calculated wave function distribution of the Fermi-arc band. g) Calculated JDOS of intra-Fermi-arc scattering processes alone. Adapted with permission.[26] Copyright 2016, American Association for the Advancement of Science.